\documentclass[pra,showpacs,preprint]{revtex4}

\usepackage{epsfig}

\newcommand{\tr}[1]{\mathop{\mathrm{tr}\left\{#1\right\}}}

\newcommand{\tfrac}[2]{{\textstyle\frac{#1}{#2}}}

\begin{document}

\title{A more efficient variant of the Oxford protocol}
\author{Nasser Metwally}
\affiliation{Sektion Physik, Universit\"at M\"unchen,
Theresienstra\ss{}e 37, 80333 M\"unchen, Germany}

\date{12 September 2001}   

\begin{abstract}
An alternative presentation  of the  Oxford purification protocol is obtained
by using dynamical variables. 
I suggest to  introduce  the degree of separability as a purification
parameter,  where the purified state has a smaller  degree of separability
than the initial one. 
An improved version  of the Oxford protocol is described,  in which local
unitary transformations optimize each step. 
\end{abstract}


\maketitle

%
Some  quantum communication proposals  require  maximally entangled qubit
pairs to perform them (see, e.g., \cite{Dirk}). 
Due to  noisy  channels, the pairs lose their fidelity partially;  
dissipative effects of the environment turn pure states into mixed states. 
The aim is then to  purify those states to re-obtain maximally entangled qubit
pairs.  
The entanglement purification that is often required  distills a small  number
of strongly entangled pairs of qubits from a larger number of weakly entangled
pairs, by using local quantum operations, classical communications, and
measurements.  

The first entanglement purification protocol, called IBM protocol has been
given by Bennett \textit{et al.} \cite{Bennett}. 
It enables   one to distill from a large ensemble of entangled states with
fidelity greater than 0.5 a smaller ensemble of pairs with fidelity close to
unity. Those purified pairs  could be used for faithful teleportation. 
Also Deutsch \textit{et al.} \cite{Deutsch} have formulated another protocol 
designed for cryptographic purposes; 
it is  called ``quantum privacy amplification'', or ``Oxford protocol'' 
for short. 

Purification under imperfect operations is studied by Giedke \textit{et al.} 
\cite{Giedke} who obtain a  lower bound for the  fidelity, such that
purification is possible in the presence of noise. 
Fidelity is the overlap of the density operator of a pair
of qubits with the wanted maximally entangled state. 

In this work the dynamical variables of the qubits are used to describe  the
Oxford  protocol, at the relevant example of the so-called Bell-diagonal 
states and their special kind known as binary states. 
The degree of separability serves as  alternative  purification
parameter, where a more entangled state has a smaller  degree of
separability. 
An improved  Oxford protocol is introduced; it converges faster and is more
efficient than the original one.

The paper is organized as follows. 
In Sec.\ \ref{sec:diagBell}, Bell-diagonal states are re-considered. 
In Sec.\ \ref{sec:OxProt}, the bilateral controlled NOT (BCNOT) 
operation is described in terms of the dynamical variables and a  description
of the  Oxford protocol is presented in   these variables. 
In Sec.\ \ref{sec:SepPur},  the degree of separability is considered as a
purity measure. 
The two variants of the Oxford protocol are investigated 
for states of two kinds, binary states and the more general 
Bell-diagonal states.
It should be noted that both variants of the Oxford protocol always produce
Bell-diagonal states after the first iteration, irrespective of whether the
initial state is of this kind or not.
Therefore, we do not need to consider more general states.

\section{Bell-diagonal states}\label{sec:diagBell}
Analogs of Pauli's spin operators are, as usual, used for the description of
the individual qubits,  the hermitian set $\sigma_x,\sigma_y,\sigma_z$ for
the first qubit and $\tau_x,\tau_y,\tau_z$ for the second. 
A ``generalized  Werner state of the first kind'' \cite{EM:JMO}, 
or ``self-transposed state'' \cite{EM:book},
or ``Bell-diagonal state'' \cite{Deutsch} is given by
\begin{equation}\label{eq:1}
\rho_{\text{Bell-diag}}
=\tfrac{1}{4}\bigl(1-c_x\sigma_x\tau_x-c_y\sigma_y\tau_y
-c_z\sigma_z\tau_z\bigr),
\end {equation}
where
\begin{equation}\label{eq:2}
1\geq\vert c_x\vert\geq\vert c_y\vert\geq\vert c_z\vert\geq 0,  
\end{equation}
the order being a matter of convention.   
This state is separable if it has a positive partial transpose \cite{PHHH}, 
which is the case if either 
$\vert c_x\vert+\vert c_y\vert+\vert c_z\vert\leq 1$ or $c_xc_yc_z\leq 0$. 
Otherwise, that is: if $\vert c_x\vert+\vert c_y\vert+\vert c_z\vert> 1$ and
$c_xc_yc_z>0$, the state is non-separable and 
\begin{equation}\label{eq:3}
{\mathcal{S}}
=\tfrac{3}{2}-\tfrac{1}{2}
\bigl(\vert c_x\vert+\vert c_y\vert +\vert c_z\vert\bigr)  
\end{equation}
is its degree of separability \cite{EM:JMO,EM:book}. 
By a suitable local unitary transformation it can then be arranged that all
$c_k$'s are positive. 
In particular, for $c_x=c_y=c_z=t$, one gets the standard Werner states
\cite{Werner},
\begin{equation}\label{eq:4}
\rho_{\text{Werner}}=\tfrac{1}{4}
\bigl[1-t(\sigma_x\tau_x+\sigma_x\tau_y+\sigma_z\tau_z)\bigr],
\end{equation}
with $-\frac{1}{3}\leq t\leq1$.
These states  are  separable for $t\leq\frac{1}{3}$ 
and non-separable for $t>\frac{1}{3}$ with the degree of separability 
given by $\mathcal{S}$=$\frac{3}{2}(1-t)$.

\section{Oxford protocol}\label{sec:OxProt}
Before performing  the Oxford protocol  on  dynamical variables, 
one needs to describe the BCNOT operation on those variables. 
In this operation, both  members of one pair are used as source qubits and
both qubits from the other pair are used as target qubits. 
The BCNOT is
\begin{eqnarray}\label{eq:5}
\mathrm{BCNOT}(\sigma^{(1)}_{\mu}\sigma^{(2)}_{\nu})&=&
\frac{1+\sigma^{(1)}_z}{2}\sigma^{(1)}_\mu\frac{1+\sigma^{(1)}_z}{2}
\sigma^{(2)}_{\nu}
\\
&&\mbox{}
+\frac{1+\sigma^{(1)}_z}{2}\sigma^{(1)}_\mu\frac{1-\sigma^{(1)}_z}{2}
\sigma^{(2)}_x\sigma^{(2)}_\nu
\nonumber\\
&&\mbox{}+
\frac{1-\sigma^{(1)}_z}{2}\sigma^{(1)}_\mu\frac{1+\sigma^{(1)}_z}{2}
\sigma^{(2)}_\nu\sigma^{(2)}_x   
\nonumber\\
&&\mbox{}+
\frac{1-\sigma^{(1)}_z}{2}\sigma^{(1)}_\mu\frac{1-\sigma^{(1)}_z}{2}
\sigma^{(2)}_x\sigma^{(2)}_\nu\sigma^{(2)}_x
\nonumber
\end{eqnarray}
where the suffixes 1 and 2 refer to the first and the second qubit. 
Table \ref{tbl:BCNOT} 
shows the effect of the BCNOT operation on the two qubits, 
that specify $\sigma^{(1)}_\mu$ and $\sigma^{(2)}_\nu$.

\begin{table}[!b]
\caption{\label{tbl:BCNOT}%
Bilateral CNOT operation between the two qubits which define 
$\sigma^{(1)}_\mu$ and $\sigma^{(2)}_\nu$. 
The same table applies for the two qubits $\tau^{(1)}_\mu$ 
and $\tau^{(2)}_\nu$, where  $\mu,\nu$=$x,y$ and $z$.}
\begin{ruledtabular}
\begin{tabular}{@{\enskip}c|cccc@{\enskip}}
&$1^{(2)}$&$\sigma_x^{(2)}$&$\sigma_y^{(2)}$ &$\sigma_z^{(2)}$\\
\hline\rule{0pt}{4ex}%
$1^{(1)}$&$1$&$\sigma^{(1)}_x\sigma^{(2)}_x$&$\sigma^{(1)}_y\sigma^{(2)}_x$
&$\sigma^{(1)}_z$ \\\rule{0pt}{4ex}%
$\sigma^{(1)}_x$&$\sigma^{(2)}_x$&$\sigma^{(1)}_x $&$\sigma^{(2)}_y$
&$\sigma^{(1)}_z\sigma^{(2)}_x$\\\rule{0pt}{4ex}%
$\sigma^{(1)}_y$&$\sigma^{(1)}_z\sigma^{(2)}_y$
&$\sigma^{(1)}_y\sigma^{(2)}_y$&$-\sigma^{(1)}_x\sigma^{(2)}_z$
&$\sigma^{(2)}_y$\\\rule{0pt}{4ex}%
$\sigma^{(1)}_z$&$\sigma^{(1)}_z\sigma^{(2)}_z$
&$-\sigma^{(1)}_y\sigma^{(2)}_y$&$\sigma^{(1)}_x\sigma^{(2)}_y$
&$\sigma^{(2)}_z$
\end{tabular}
\end{ruledtabular}
\end{table}

In this protocol the users Alice and Bob have a supply of qubit pairs, each
pair being in the pure, maximally entangled state,  
\begin{equation}\label{eq:6}
\rho_{\text{ideal}}=\frac{1}{4}(1+\sigma_x\tau_x-\sigma_y\tau_y+\sigma_z\tau_z).
\end{equation}
Because of the noise along the transmission channel, the pairs interact with
the environment, so they lose their purity.  
Assume that Alice and Bob are given an ensemble that consists of two
subensembles. 
Each of those subensembles  is made of Bell-diagonal states with different
$c_k$'s. 
Let Alice and Bob  pick two different pairs, one from each subensemble,
\begin{eqnarray}\label{eq:7}
\rho^{(1)}&=&\tfrac{1}{4}\bigl(1+c_{x}\sigma_x^{(1)}\tau_x^{(1)}-c_{y}
\sigma_y^{(1)}\tau_y^{(1)}+c_{z}\sigma_z^{(1)}\tau_z^{(1)}\bigr), 
\nonumber\\
\rho^{(2)}&=&\tfrac{1}{4}\bigl(1+ c_{x}^\prime\sigma_x^{(2)}\tau_x^{(2)}
- c_{y}^\prime\sigma_y^{(2)}\tau_y^{(2)}+
c_{z}^\prime\sigma_z^{(2)}\tau_z^{(2)}\bigr),
\end{eqnarray}
with  fidelities
\begin{eqnarray}
F_1&=&\tr{\rho^{(1)}\rho^{(1)}_{\text{ideal}}}
=\tfrac{1}{4}(1+ c_{x}+ c_{y}+ c_{z}),
\nonumber\\
F_2&=&\tr{\rho^{(2)}\rho^{(2)}_{\text{ideal}}}
=\tfrac{1}{4}(1+ c_{x}^\prime+ c_{y}^\prime+c_{z}^\prime).
\end{eqnarray}

In the original protocol, ${\text{Ox}}_1$, Alice and Bob perform the
transformation $U_{12x}=e^{i\pi(\sigma_x-\tau_x)/4}$ on all pairs. 
This operator changes the positions of $c_y$ and $c_z$ in (\ref{eq:7}). 
Then  Alice and Bob perform  BCNOT operations  on the pairs $\rho^{(1)}$
and $\rho^{(2)}$, followed by measuring the target qubits in the computational
basis. 
For example,  they measure the $z$ components of the targets spin,
$\sigma^{(2)}_z$ and $\tau_z^{(2)}$. 
They keep those first pairs for which they get the same measurement results,
and discarded the others. 
The target pairs are always consumed in the process. 

In the alternative protocol, ${\text{Ox}}_2$, 
one exploits the order specified in (\ref{eq:2}) 
and performs BCNOT directly, without first applying $U_{12x}$. 
The resulting subensemble of good first pairs is characterized by
\begin{eqnarray}\label{eq:9}
\rho_{\text{new}}&=&
\frac{1}{4}
\biggl[1+\frac{c_{x} c_{x}'+c_{y} c_{y}'}
{1+c_{z} c_{z}'}\sigma_x\tau_x-\frac{c_{z} c_{z}'
+c_{y} c_{x}'}{1+c_{z} c_{z}'}\sigma_y\tau_y
\nonumber\\
&&\hspace*{1.9 em}
\mbox{}+\frac{c_{z}+ c_{z}'}{1+c_{z} c_{z}'}\sigma_z\tau_z\biggr].
\end{eqnarray}
This is another Bell-diagonal state.

In the standard description of ${\text{Ox}}_1$ \cite{Deutsch,Dirk}, 
certain parameters  $A$, $B$, $C$, and $D$ play a central role. 
Their change under ${\text{Ox}}_2$ is given by
\begin{eqnarray}\label{eq:10}
A&=&\left\{
  \begin{array}{c}
\frac{1}{4}(1+ c_{x}+ c_{y}+ c_{z})\\[1ex]
\frac{1}{4}(1+ c'_{x}+ c'_{y}+ c'_{z})
  \end{array}
\right\}
\nonumber\\
&\to&\frac{1}{4N}\Bigl[(1+c_z)(1+ c_z')+(c_x+c_y)( c_x'+c_y')\Bigr],
\end{eqnarray}
for example, and corresponding expressions apply for $B$, $C$, and $D$.
Here $N= \frac{1}{2}(1+ c_z c_z')$ is the probability that Alice and Bob
obtain coinciding outcomes in the measurements of the target pair. 
If one changes the positions of $c_y$ and $c_z$ and also of $c_y'$ and $c_z'$
in (\ref{eq:10}), one gets the $A$, $B$, $C$ and $D$ values for 
${\text{Ox}}_1$.
 
If the two subensembles in (\ref{eq:7}) are  identical,  
then the protocol works if  $F_1=F_2>\frac{1}{2}$. 
In terms of the parameters of (\ref{eq:7}), this means 
\begin{equation}\label{eq:11}
\vert c_x\vert+\vert c_y\vert+\vert c_z\vert>1.
\end{equation} 
So, at every step Alice and Bob must check this property. 
In particular,  they need 
\begin{equation}\label{eq:12}
(\vert c_x\vert+\vert c_y\vert)^2-(1-\vert c_z\vert)^2 >0
\end{equation}
for the first step to be successful.

If the given ensemble  does not obey the ordering required by (\ref{eq:2}), 
then Alice and Bob use  unilateral rotations to bring the state into the
wanted form.   
These are rotations by $\pi$ about the  $x,y$ or $z$ axis, namely
\begin{eqnarray}
U_{1x}&=&\sigma_x,\quad U_{1y}=\sigma_y,\quad U_{1z}=\sigma_z,
\nonumber\\
U_{2x}&=&\tau_ x,\quad U_{2y}=\tau_y,\quad U_{2z}=\tau_z,
\label{eq:13}
\end{eqnarray}
where $U_1$ and $U_2$ refer to the first and second qubit, respectively.
In fact, it is only necessary to ensure that $|c_z|$ is smaller than $|c_x|$
and  $|c_y|$; the relative size of $|c_x|$ and  $|c_y|$ does not matter.

\section{Separability and purification}\label{sec:SepPur}
In this section  the degree of separability is used as a purification
parameter instead of the fidelity. 
Also, the behavior of the degree of separability under imperfect operations is
investigated. 
Two cases are considered: Binary states and the more general 
Bell-diagonal states.  

\textbf{(1) Binary state with perfect operations:}
In this case,
\begin{equation}\label{eq:14}
\rho_{\text{bin}}=\tfrac{1}{4}\bigl[1+\sigma_x\tau_x-(2f-1)\sigma_y\tau_y
+(2f-1)\sigma_z\tau_z\bigr],
\end{equation}
with the initial degree of separability
\begin{equation}\label{eq:15}
{\mathcal{S}}_0=
\left\{\begin{array}{cl}
1&\mbox{for $0<f\leq\frac{1}{2}$},\\[1ex]
2(1-f)&\mbox{for $\frac{1}{2}<f<1$}.
\end{array}\right.
\end{equation}
Assume that Alice and Bob  are given an ensemble of states (\ref{eq:14}), 
and they are  asked to  purify this ensemble. 
They perform the ${\text{Ox}}_2$  protocol and after one step they get
\begin{eqnarray}
\rho_{\text{bin}}'&=&
\frac{1}{4}\biggl[1+\sigma_x\tau_x-\frac{2f-1}{2f^2-2f+1}\sigma_y\tau_y
\nonumber\\
&&\hspace*{2.1 em}
\mbox{}+\frac{2f-1}{2f^2-2f+1}\sigma_z\tau_z\biggr].
\label{eq:16}
\end{eqnarray}
The corresponding  degree of separability is
\begin{equation}\label{eq:17}
{\mathcal{S}}_1=\frac{{\mathcal{S}}^2_0}{1+(1-{\mathcal{S}}_0)^2}.
\end{equation}
After repeating the protocol $n$ times one gets
${\mathcal{S}}_n$  as a function of the initial 
degree of separability ${\mathcal{S}}_0$,
\begin{equation}\label{eq:18}
{\mathcal{S}}_n=\frac{2}{\left(2/{\mathcal{S}}_0-1\right)^{2^{n}}+1}.
\end{equation}
From this relation it is clear that  ${\mathcal{S}}_n=1$  
if  ${\mathcal{S}}_0=1$ and  ${\mathcal{S}}_n\to 0$ if ${\mathcal{S}}_0<1$. 

\textbf{(2) Binary state with imperfect operations:}
In this case the operations  are subjected to  noise, so that states of two
qubit pairs suffer a non-unitary evolution such that \cite{Dur} 
\begin{equation}\label{eq:19}
\rho_{12}\to p\rho_{12}+(1-p)\tfrac{1}{2}\mathrm{tr}_{1}\{\rho_{12}\},
\end{equation}
where $p$ is called reliability of the imperfect operation. 
The limit $p\to 0$ corresponds to a  very noisy channel, while $p\to 1$
describes a channel with  very little noise. 
For two pairs in the binary state (\ref{eq:14}), the map (\ref{eq:19}) produces
\begin{equation}\label{eq:20}
\rho_{\text{bin}}^{\text{noise}}
=\tfrac{1}{4}\bigl[1+p\sigma_x\tau_x-p(2f-1)\sigma_y\tau_y
+p(2f-1)\sigma_z\tau_z\bigr]
\end{equation}
for the ``first'' pairs.
Rather than (\ref{eq:15}) the initial degree of separability is now
\begin{equation}\label{eq:21}
{\mathcal{S}}_0=\tfrac{1}{2}\bigl[3-p(4f-1)\bigr].
\end{equation}
Further, the ideal BCNOT operation of (\ref{eq:5}) 
is replaced by ${\text{BCNOT}}_{\text{noise}}$,
\begin{equation}\label{eq:22}
\mathrm{BCNOT}_{\mathrm{noise}}(.)=p^2\,\mathrm{BCNOT}(.)+\frac{1-p^2}{16},
\end{equation}
where $(.)$ is $\rho^{(1)}\rho^{(2)}$. 
Alice and Bob perform the ${\text{Ox}}_2$  protocol, 
and  after the measurement of the target qubits and discarding of 
the ``bad'' first pairs they obtain
\begin{eqnarray}
\rho&=&\frac{1}{4}\biggl[1+p^2\frac{2f^2-2f+1}{1-2p^2f(1-f)}\sigma_x\tau_x
\nonumber\\
&&\hspace*{1.9 em}
\mbox{}-p^2\frac{2f-1}{1-2p^2f(1-f)}\sigma_y\tau_y
\nonumber\\
&&\hspace*{1.9 em}
\mbox{}+p^2\frac{2f-1}{1-2p^2f(1-f)}\sigma_z\tau_z\biggr],
\label{eq:23}
\end{eqnarray}
for the ``good'' first pairs.
The new degree of separability is  
\begin{equation}\label{eq:24}
{\mathcal{S}}_{\text{new}}
=\frac{1}{2}\left[3-p^2\frac{2f^2+2f-1}{1-2p^2f(1-f)}\right].
\end{equation}

\textbf{(3) Bell-diagonal state:}
Now consider the ensemble (\ref{eq:7}) consisting  of  Bell-diagonal states. 
In this case the initial degrees of separability  are given by
\begin{eqnarray}
{\mathcal{S}}_0&=&\tfrac{3}{2}-\tfrac{1}{2}
\bigl(\vert c_x\vert +\vert c_y\vert+\vert c_z\vert\bigr),
\nonumber\\
{\mathcal{S}}'_0&=&\tfrac{3}{2}-\tfrac{1}{2}
\bigl(\vert c'_x\vert+\vert c'_y\vert+\vert c'_z\vert\bigr).
\end{eqnarray}
Alice and Bob perform the ${\text{Ox}}_2$  protocol, 
and after one step they get
\begin{equation}
{\mathcal{S}}_1=\frac{3}{2}-\frac{1}{2N}
\bigl[\bigl(\vert c_x\vert+\vert c_y\vert\bigr)
\bigl(\vert c'_x\vert+\vert c'_y\vert\bigr)
+\vert c_z\vert+\vert c'_z\vert\bigr]
\end{equation}
for the ``good'' first pairs with $N$ as in (\ref{eq:10}), 
or in the  presence  of  noise,
\begin{equation}
N_{\text{noise}}
=\frac{1}{4p^2}\bigl[1+p^2\bigl(1+2\vert c_z c_z'\vert\bigr)\bigr].
\end{equation}

\begin{figure}[!t]
\epsfig{file=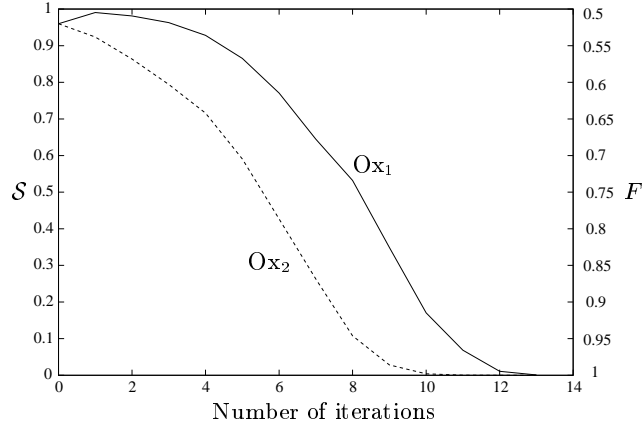}
\caption{\label{fig:1}%
The degree of separability $\mathcal{S}$ and fidelity $F$ 
for the two variants of the Oxford protocol. 
Solid line: original protocol ${\text{Ox}}_1$; 
dashed line:  alternative protocol   ${\text{Ox}}_2$ .} 
\end{figure}

\begin{figure}
\epsfig{file=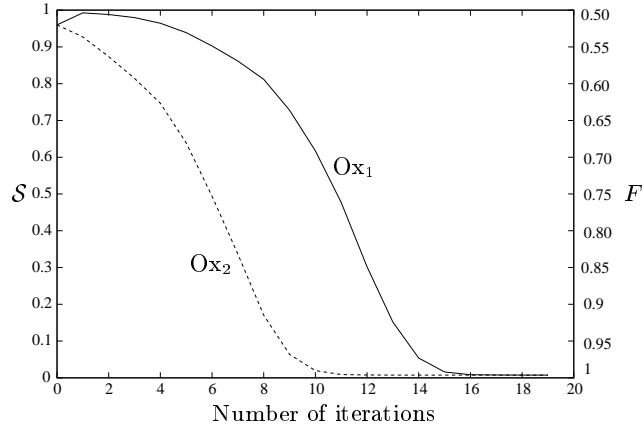}
\caption{\label{fig:2}%
Like Fig.\ \ref{fig:1}, but with noise of strength $p=0.994$.} 
\end{figure}

\begin{figure}
\epsfig{bbllx=60,bblly=571,bburx=295,bbury=740,clip=,
file=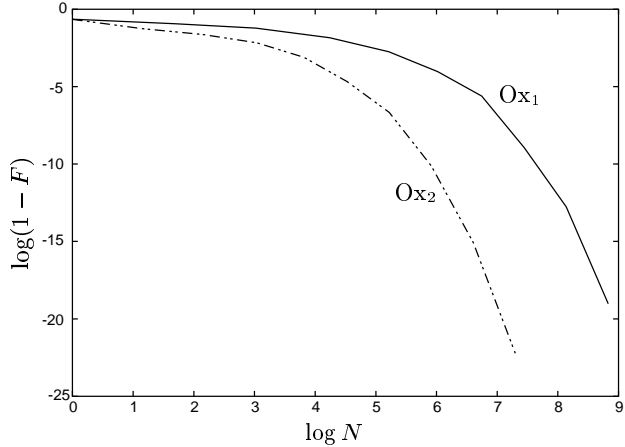}
\caption{\label{fig:3}%
Number $N$ of pairs needed to create one pair with fidelity $F$, 
displayed as $\log(1-F)$ vs.\ $\log N$.
The initial state of the pairs has fidelity $F_0=0.62$.}
\end{figure}

\section{Discussion and Conclusion}
In Fig.\ \ref{fig:1}, the separability $\mathcal{S}$ and the fidelity $F$ are
plotted as a function of the number of iterations, both  one for the original
protocol $\mathrm{ Ox_1}$ and  for the alternative protocol, ${\text{Ox}}_2$,
where  one enforces the ordering of (\ref{eq:2}) in each step. 
The figure refers to  the initial  values  $( c_x, c_y, c_z)=(0.16,0.08,0.84)$
for which $F=0.52$ is the initial fidelity and $2F+\mathcal{S}$=$2$ holds for
all iterations. 
In this case Alice and Bob use the bilateral rotations to rearrange these
three numbers  such that $(c_x, c_y, c_z)=(0.84,0.16,0.08)$. 
The figure clearly shows that for ${\text{Ox}}_2$, 
the fidelity reaches  unity much faster than that for ${\text{Ox}}_1$. 
Moreover, for ${\text{Ox}}_1$ the fidelity decreases and then increases
\cite{Chiara}, but for ${\text{Ox}}_2$ it increases in each iteration. 

The importance of (\ref{eq:2}) is particularly apparent when one treats the
binary state (\ref{eq:14}), for which two of the $A, B, C, D$ parameters of 
\cite{Deutsch,Dirk} are positive and the other two vanish. 
To perform ${\text{Ox}}_1$ successfully, one needs $A,C>0$ and $B=D=0$; 
then ${\text{Ox}}_1$ works and the fidelity increases monotonically. 
But if one enforces (\ref{eq:2}), ${\text{Ox}}_2$ works directly, 
and one doers not have to worry which of the four parameters are non-zero.

In Fig.\ \ref{fig:2}, ${\text{Ox}}_1$ and ${\text{Ox}}_2$ 
are performed in the presence of  noise. 
The importance of the property (\ref{eq:2}) is  clear: 
the degree of separability  becomes constant faster for ${\text{Ox}}_2$ 
than for ${\text{Ox}}_1$.

The log-log plot of Fig.\ \ref{fig:3} shows the number of initial pairs 
needed to create one pair with fidelity $F$.
We see that ${\text{Ox}}_2$ uses up less qubit pairs than ${\text{Ox}}_1$. 
In addition, ${\text{Ox}}_2$ needs fewer iterations, so that both advantages
taken together make ${\text{Ox}}_2$ much more efficient than ${\text{Ox}}_1$.

In summary, in this contribution an alternative form of the  Oxford protocol
is described for Bell-diagonal states. 
The final fidelity is obtained as a function of three numbers, 
$c_x$, $c_y$ and $c_z$. 
The  improvement over the original Oxford protocol is  due to the
arrangement of these three numbers in decreasing order. 
The parameter of the degree of separability is considered as a purification
parameter.  
As the number of iterations increases, the degree of separability decreases.

\begin{acknowledgments}
I am grateful to B.-G. Englert and H.-J. Briegel  for their useful
discussions. Also my thanks to H. Aschauer and R. Raussendorf for their
important remarks. This work is supported by a Grant from the Egyptian
Embassy.
\end{acknowledgments}

\newcommand{\PRL}[3]{Phys.\ Rev.\ Lett.\ \textbf{#1}, #2 (#3)}
\newcommand{\PRA}[3]{Phys.\ Rev.\ A \textbf{#1}, #2 (#3)}
\newcommand{\plA}[3]{Phys.\ Lett.\ \textbf{A#1}, #2 (#3)}
\newcommand{\JMO}[3]{J. Mod.\ Opt.\ \textbf{#1}, #2 (#3)}

\end{document}